\documentclass[10pt]{article} 
\usepackage{times}
\usepackage{graphicx}
\usepackage{amssymb}
\usepackage{url,hyperref}
\usepackage{multicol}

\begin{document}

\title{Evaluating District-based Election Surveys with Synthetic Dirichlet Likelihood}

\author{Adway Mitra, Palash Dey\\
\\
Indian Institute of Technology Kharagpur\\
}

\maketitle

\begin{abstract}
In district-based multi-party elections, electors cast votes in their respective districts. In each district, the party with maximum votes wins the corresponding “seat” in the governing body. Election Surveys try to predict the election outcome (vote shares and seat shares of parties) by querying a random sample of electors. However, the survey results are often inconsistent with the actual results, which could be due to multiple reasons. The aim of this work is to estimate a posterior distribution over the possible outcomes of the election, given one or more survey results. This is achieved using a prior distribution over vote shares, election models to simulate the complete election from the vote share, and survey models to simulate survey results from a complete election. The desired posterior distribution over the space of possible outcomes is constructed using Synthetic Dirichlet Likelihoods, whose parameters are estimated from Monte Carlo sampling of elections using the election models. We further show the same approach can also use be used to evaluate the surveys - whether they were biased or not, based on the true outcome once it is known. Our work offers the first-ever probabilistic model to analyze district-based election surveys. We illustrate our approach with extensive experiments on real and simulated data of district-based political elections in India.
\end{abstract}

\section{INTRODUCTION}
Elections are conducted by almost all democratic countries to choose representatives for governing bodies, such as parliaments. A common democratic setup is the district-based system in which the country is spatially divided into a number of regions called districts (or constituencies). There is a seat in the governing body corresponding to each district. The residents of each district elect a representative from a set of candidates, according to any voting rule. In many countries, these candidates are representatives of political parties, and electors may cast their votes in favour of the parties rather than individual candidates. 

The election results are understood in terms of the number of seats won by different parties, rather than the total number of votes obtained by them. If the relative popularity of the different parties is spatially homogeneous across all the districts, then the most popular party may win all the seats. But this is very rarely the case. One reason for this may be the individual popularity of candidates may vary. But a more complex reason is the spatial variation of demography across the country, since the popularity of different parties often varies with demography~\cite{b}. Demography varies spatially as people usually prefer to choose residences based on social identities, such as race, religion, language, caste, profession and economic status. This process is sometimes called ``ghettoization", where people with similar social identities huddle together in pockets~\cite{d,e}. Such ghettoization plays a very important role in district-based elections if different political parties represent the interests of different social groups. Even if a political party is not popular overall, it can win a few seats if its supporters are densely concentrated in a small number of districts, which forms strongholds of the party. On the other hand, a party which is overall quite popular, may fail to win many seats if its supporters are spread all over without concentration. Also, electors often vote according to the advice of local community leaders and other local factors~\cite{c}, which causes ``polarization" of voters in favour of one/two parties inside each district. 

Surveys are often carried out to forecast the election results. These surveys may be conducted by various agencies before or after the election. Usually a survey involves a small sample of the electorate, based on whose responses the vote share of the different parties is estimated. The number of seats to be won by the different parties can be estimated as well from this sample. However, the accuracy of these estimates depends on how well these samples represent the entire population. For example, the chosen samples may cover only a few districts, or misrepresent the true vote share of the different parties. This may arise either due to practical constraints (such as the difficulty of reaching certain geographical areas) or due to malicious intent or partisan bias of the survey agency. This gives rise to two complementary questions: i) Given a survey method and results, can we predict the true results of the election? ii) Once the full results of the election are known, can we figure out if the estimated result from any survey is consistent with a particular survey method?

A significant amount of research work exists in predicting the election results from a survey under different conditions. Most of these works like~\cite{pred1,perse2016media,dwi2015twitter,leigh2006competing,kennedy2017improving} focus on finding the minimum number of samples needed by a survey to forecast the winner and/or the margin of victory with a given confidence, and efficient algorithms for the same. ~\cite{distpred} extends this analysis to district-based settings, and provides algorithms to carry out the survey over a limited number of districts and a limited number of persons in each district. However, none of these works, to the best of our knowledge, predict the number of districts won by the parties in either deterministic or probabilistic way. Nor are we aware of any attempt to evaluate if a given survey result is consistent with the actual results.

The aim of this work is threefold. First of all, we attempt to provide a probability distribution over the space of all possible results, given a particular survey result and its various parameters. Here, an election result indicates both the vote share and seat share of different parties. Secondly, given the actual results, we attempt to provide a distribution over the space of possible survey results. This in turn can be used to check whether a given survey result is conceivable or not. Our final aim is to evaluate the above for actual district-based elections held in India. 

Our approach depends heavily on the simulation of election outcomes. There are relatively few statistical models for this purpose. Eggenberger and Polya used the concept of Polya's urn to propose a statistical voting model, which simulates the effect that if one candidate gets a vote, there are likely to get more~\cite{berg1994}. There have been attempts to extend these to multiple districts ~\cite{multisim}. Another popular approach is Mallow's Model, which assumes a `central' ranking over the candidates, and simulates individual votes by perturbing it. More recently, there have been attempts to systematically represent various aspects of district-based elections through voter-centric agent-based statistical models~\cite{mitra2021electoral,mitra2023agent}. In this work, we utilize some of these models to simulate complete election results.

The main contribution of the work is to cast the problem in a Bayesian setting by defining conditional distribution of the actual outcome given the survey, and vice versa. These are modelled as Dirichlet Distributions, whose parameters can be estimated from samples of election surveys, drawn from complete election outcomes. Our second contribution is a probabilistic model for surveys, based on complete election outcome. Our third contribution is to propose an algorithm based on Approximate Bayesian Computation to identify the modal (most likely) outcomes, given a survey result. Next, we show how the above framework can be used to evaluate survey results using actual outcomes, to test whether they are feasible and consistent with the uniform sampling paradigm. Finally, we validate this approach through extensive experiments over both simulated and real data. This involves political elections in India covering millions of voters and multiple parties. The novelty of the work lies in the aims, approach and the empirical analysis.

\section{Notations and Problem Definition}
We consider district-based 1-plurality elections, i.e. the candidate/party with maximum votes in a district wins the corresponding seat. Consider $N$ voters divided among $S$ districts as $\{N_1,\dots,N_S\}$. There are $K$ parties in fray, each of whom has a candidate in each district. Denote by $\theta_{sk}$ the votes received by party $k$ in district $s$, and by $\theta_k$ its overall vote. Also denote by $V_k$ the number of districts where the candidate from party $k$ is the winner with maximum number of votes. Clearly, $\sum_k\theta_k=N$ and $\sum_kV_k=S$. 

Denote by $X$: the actual electoral outcome. It has two parts: $X=\{X_1,X_2\}$ where $X_1=\{\frac{\theta_1}{N},\dots,\frac{\theta_K}{N}\}$,  and $X_2=\{\frac{V_1}{S},\dots,\frac{V_K}{S}\}$ i.e. the vote shares and seat shares of the parties. Denote by $Y$: the projected results based on the surveys, which also has two parts: $\{Y_1,Y_2\}$ which are the projected vote shares and seat shares of all the parties.

Denote by $Z$ the complete election, where $Z=\{Z_1,Z_2,\dots,Z_S\}$ where $Z_s=\{\theta_{s1},\dots,\theta_{sK}\}$ denotes the vote share of the parties in district $s$. Note that the overall vote share and seat share of all parties can be easily calculated given $Z$. An election simulation model generates $Z$ given $X_1$ (note that $X_2$ can be calculated easily from $Z$). A survey model simulates $Y$ from $Z$. 

The first task is: given a set of $M$ surveys $y^1,\dots,y^M$, calculate a posterior distribution $p(X|\{y^1,\dots y^M\})$, at least till a proportionality constant. Even if the normalization factor cannot be calculated, we should still be able to compare different candidate outcomes. A related aim is to estimate the mode $argmax_X p(X|\{y^1,\dots y^m\})$, i.e. the most likely outcome.

The second task is the reverse: given the results $x$, calculate the distribution $p(Y|x)$. This shows how likely is a survey (done under certain conditions) to produce a particular projection. If the projected result of a survey (claimed to have been done under the same conditions) has very low density under this distribution, then we can doubt about its actual methodology.

\section{Model}
Now, we describe the model in full details. This has three building blocks: the posterior construction using Dirichlet synthetic likelihood, the election simulation models and the survey models. Below, we discuss each of these aspects in details.

\subsection{Constructing the Posterior}

Our main aim is to model the probability distribution $p(X|Y)$ over possible outcomes $X$, given survey projection results $Y$. Using the Bayes Theorem, we can write $p(X|Y)\propto q(Y|X)r(X)$. 

The prior $r(X)$ on $X$ can be written as $r(X)=g(X_1)*f(X_2|X_1)$. Since $X_1$ satisfies the definition of a PMF (vote proportion of the $K$ parties), it is intuitive to use the Dirichlet distribution here. So we write $g(X_1)=Dir(\gamma_1,\dots,\gamma_K)$, where $(\gamma_1,\dots,\gamma_K)$ are hyper-parameters that indicate our prior beliefs about the relative popularity of the different parties (maybe based on past elections). 

Now we introduce the complete election $Z$ through an election model which represents $h(Z|X_1)$ and survey model, which represents $q(Y|Z)$. Using them, we can write the posterior as follows: 
\begin{equation}\label{eq:model1}
p(X|Y)\propto \int_Zq(Y|Z)f(X_2|Z)h(Z|X_1)g(X_1) 
\end{equation}

Note that $f(X_2|Z)$ is deterministic, i.e. if we known the complete election result, then we can easily calculate the number of seats won by the parties. Now, both the election model and the survey model are simulation-based, i.e. we can sample $Z$ given $X_1$ and $Y$ given $Z$ respectively, but we have no analytical representation for $q$ and $r$. So the integration is intractable, and hence we need to use Approximate Bayesian Computation based on Monte Carlo Sampling, as follows:
\begin{equation}\label{eq:model2}
p(X|Y)\propto \frac{1}{M}\sum_{i=1}^Mq(Y|Z_i)f(X_2|Z_i)g(X_1)    
\end{equation} 
where $Z_i$ are sampled from the election model $h(Z|X_1)$. 

Note that $Y$ has two parts $\{Y_1,Y_2\}$, the vote share and the seat share of the parties. In the absence of a theoretical representation of $q(Y|Z)$, we can consider \emph{Synthetic Likelihood} for them, like several works on Approximate Bayesian Inference~\cite{syn1, syn2}. As both of them are proportions, Dirichlet Distribution is a sensible choice for such synthetic likelihood. The parameters $\alpha=\{\alpha_1,\dots,\alpha_K\}$ and $\beta=\{\beta_1,\dots,\beta_K\}$ of these distributions need to be estimated, based on samples of $Z$. 
\begin{equation}
q(Y|Z) = Dir(Y_1|\alpha(Z))*Dir(Y_2|\beta(Z))
\end{equation}
We can write this because given $Z$, $Y_1$ and $Y_2$ can be considered as conditionally independent. This is ensured by the way that the survey model works. Here $\alpha(Z_i), \beta(Z_i)$ are complex functions of $Z_i$. One possibility might be to represent them using Neural Networks, but here we again use another Monte Carlo approach:
\begin{eqnarray}
&\alpha(Z_i)& = \textit{argmax}_{\alpha}\prod_{j=1}^L Dir(y_{1j}|\alpha) \textit{ and} \nonumber \\
&\beta(Z_i)& = \textit{argmax}_{\beta}\prod_{j=1}^L Dir(y_{2j}|\beta) \nonumber \\
\textit{ where} &y_j& \sim q(y_j|Z_i)
\end{eqnarray}

Here, $y_j$ are $L$ sample surveys drawn from the true election $Z_i$ according to the survey model $q$. Estimated vote shares $y_{1j}$ and seat shares $y_{2j}$ are obtained from them. Our synthetic Dirichlet likelihood is applicable for them too. Using these samples, maximum-likelihood estimates of $(\alpha,\beta)$ are calculated, using the algorithms discussed in~\cite{dirich}. These ML estimates are used to calculate the likelihood of the actual survey $Y$, using the synthetic Dirichlet likelihood again.

\subsection{Election Models}
Suppose we know the total number of voters in support of the different parties. However, the outcome of the election is unknown, as it depends on how these voters are distributed across the districts. To take a small example, let us consider two parties A and B, which have 15 and 10 supporters respectively. These 25 voters are spread over 5 districts, each of which have 5 voters.Now if the spread is uniform, i.e. each district has 3 voters for party A and 2 voters for party B, then party A wins all 5 districts. On the other hand, if all voters in 3 districts support A while all voters in the other 2 districts support B, then A wins 3 districts and B wins 2. But if two districts have only A voters, while the remaining 5 A voters are spread across the remaining 3 districts as (2,2,1), then party A wins only the first 2 districts, while party B wins the remaining 3 districts despite having less supporters. To explore the space of possible electoral outcomes, it is thus necessary to consider different possible spatial distributions of the voters, given the overall popularities of the parties $\{\theta_1,\dots,\theta_K\}$. The aim of the election model is to achieve this through sampling. 

While simulating the spread of voters across districts, it is necessary to make sure that these distribution patterns are realistic. Real-world political elections have certain characteristics, such as i) In a district, most of the voters support a small subset of parties in fray, ii) People supporting any party are more likely to be staying in the same districts. These happen due to various sociological factors that influence electoral preferences, especially in a heterogeneous society where political preferences often depend on social identity. An Election Model should be able to produce these features in its simulation. 

One of the most well-known election simulation models that partially captures the first aspect mentioned above is 
the Polya Urn model, which works on the idea that if one voter chooses a candidate, then the probability of subsequent voters choosing the same candidate increases. However, this is restricted to the single-district case. 
We consider the agent-based models proposed in~\cite{mitra2021electoral} for district-based elections. 
These models focus on each voter as an agent, and assign them to a district and/or party according to a probabilistic process to maintain the above two properties.

We first consider the Districtwise/Seatwise Polarization Model (SPM) that has a single parameter $\gamma$, called \emph{concentration parameter}. The idea is based on Chinese Restaurant Process~\cite{crp} similar to Polya's Urn. Each voter in a district is likely to choose a party according to its local popularity (number of votes it has already received in same district) with probability $\gamma$, while with the remaining probability $1-\gamma$ they can choose a party according to the overall popularity. In general, high value of $\gamma$ causes concentration of support of parties in specific districts, so that the seat share is a reflection of the overall popularities of the parties. On the other hand, low value of $\gamma$ causes the vote share in each district to reflect the overall popularities (vote shares) of the parties, and thus the most popular party wins almost all the seats.

It often happens that a party with high vote share wins fewer seats than a less popular party, because its voters are either too concentrated (reducing spatial spread) or too diffuse (failing to achieve adequate concentration to win any district). This phenomena cannot be captured by the SPM. So we consider the Partywise Concentration Model (PCM) with party-specific concentration parameters $\{\gamma_1,\dots,\gamma_K\}$. This model places each voter in a district which already has other voters who support the same party $k$, with probability $\gamma_k$\. However, with probability $1-\gamma_k$, the voter is placed in any district uniformly. Different combinations of high/low values of these party-specific parameters can create widely differing and unexpected results. The PCM model is much richer than SPM as it can simulate a much broader spectrum of results, but is also more difficult to calibrate as it as $K$ parameters.

\subsection{Survey Models}
The aim of a survey is to estimate the underlying reality by examining a small number of samples. In this case, the underlying reality is the actual voting preference of all voters, i.e. $Z$, and the aim of the survey is to predict the vote shares $X_1$ and seat shares $X_2$. This is obtained by selecting a small subset of the voters and finding out their preferences (it is assumed that they respond truthfully). 

The main question here is, how to choose these respondents. As already discussed, the preferences may vary from district to district. While it may not be possible to cover all districts, an unbiased survey can be considered to choose districts uniformly at random, and also choose respondents uniformly at random from these districts. This approach of uniform sampling has been discussed by other works like ~\cite{distpred}, which provided lower bounds on the fraction of districts to be sampled, and the number of people to be queried in each district to be able to predict the winner correctly. In our model, we leave these as parameters $f_s$ and $f_n$. We further assume that equal number of people are queried in all the chosen districts.

Suppose in district $j$, we find $\{W_{j1},\dots,W_{jK}\}$ respondents in favour of the $K$ parties. Clearly, this follows a Multinomial Distribution with parameters $\{N_jf_n,(\theta_{j1},\dots,\theta_{jK})\}$. The next question is, given the survey results, how to predict the outcome $\{X_1,X_2\}$. Our model estimates the total vote share by simply aggregating the number of respondents across all districts, who expressed preferences for different parties. In other words, $Y_1(k)=\frac{\sum_jW_{jk}}{Nf_n}$ ($Nf_n$ is the total number of respondents) for party $k$. Next, in each of the $Sf_s$ districts where we carried out the survey, we identify the party with maximum number of votes among the respondents from that district. Thus, we find the number of districts $\{v_1,\dots,v_K\}$ ``won" by the different parties, and we use this as our estimate $Y_2$ of the overall seat share, i.e. $Y_2(k)=\frac{v_k}{Sf_s}$.

\section{Analysis of Elections}
As already discussed, our aims in this paper are twofold- prediction of the results based on the surveys, and evaluating the surveys based on the results. We now discuss how these can be achieve these using the model discussed above.

\subsection{Prediction from Surveys}
Consider the situation where $M$ surveys have been conducted, with results $\{y_1,\dots,y_M\}$, where $y_i=\{y_{i1},y_{i2}\}$ and we aim to estimate $X$ from them. We have already described our approach to construct the posterior $p(X|y_1,\dots,y_m)$. However, this construction does not account for the normalization factor $\frac{1}{p(Y)}$. Even if it were known, it would be difficult to visualize the infinite space of possible outcomes. 

We discuss two ways to utilize this posterior on possible outcomes. The first one is comparison of a finite number of candidate outcomes. We are often interested in very specific questions like, how many votes a particular party may win, or which party can win maximum seats, rather than the exact vote and seat shares of all parties. Accordingly, we can construct a few representative outcomes $x_1, \dots, x_k$, and compare their relative likelihoods through $p(x_i|y_1,\dots,y_m)$.

Also, often the seat share is more important than the vote share, and there are only a finite number of seat shares (based on how $S$ seats can be distributed among $K$ parties). So a PMF can be constructed by calculating the posterior measure for each possible seat share, and normalizing them.

If we need a distribution for an individual party's vote share or seat share, it is difficult to calculate it analytically from the above model, because the constructed posterior does not follow a known family of distributions. However, we can still use a Monte Carlo approach again if we can draw samples from an approximate form of the posterior. The proposed approach is as follows:
\begin{enumerate}
    \item Initialize sample set $\mathcal{S}=\Phi$
    \item Draw a sample $x_1$ from prior $r$
    \item Simulate an election $z$ based on $x_1$ using Election Model
    \item Calculate $x_2$ from $z$
    \item Simulate a survey $y$ from $z$ using Survey Model
    \item If $y$ is close enough to the observed surveys $\{y_1,\dots,y_M\}$, ACCEPT the sample, else REJECT it
    \item If sample is ACCEPTED, add $\{x_1,x_2\}$ to $\mathcal{S}$
    \item Repeat till we have sufficient samples
\end{enumerate}
Step 6 ensures that the accepted samples are consistent with the surveys. Any suitable measure like Kullback-Leibler (K-L) divergence can be used to compare $y$ with $\{y_1,\dots,y_M\}$. The ranks of the different parties with respect to the different estimates should also be compared.  

Once we have enough samples of $X|\{y_1,\dots,y_M\}$, we can fit another synthetic likelihood on $X$. Once again, we use Dirichlet likelihood as $X_1$, $X_2$ are both proportions over $K$ parties. Once again, the parameters $\gamma=\{\gamma_1,\dots,\gamma_K\}$ and $\eta=\{\eta_1,\dots,\eta_K\}$ can be estimated using~\cite{dirich}.
The marginal distribution of each variate in a Dirichlet distribution follows a Beta distribution.
Using this property, we can easily calculate the marginal distribution over the vote-share and seat-share of any party $k$, as follows:
\begin{equation}
X_{1k}\sim Beta(\gamma_k, \sum_{j=1}^K\gamma_j-\gamma_k), X_{2k}\sim Beta(\eta_k, \sum_{j=1}^K\eta_j-\eta_k)
\end{equation}

\subsection{Investigating the Surveys}
An election survey is supposed to be uniform and unbiased. Once the election result $x$ is known, we want to verify if the reported survey result $y$ was consistent with it. In other words, is the probability $p(Y=y|X=x)$ high enough, if the uniform survey approach was indeed followed? If not, the survey result may be considered as dubious.

We have already discussed the use of synthetic Dirichlet likelihood for $q(Y|Z)$. Given the observations $x$, we generate many samples of $Z$ (the complete election) using the Election Model, generate projected result $Y$ for each of them using the Survey Model, and then estimate the Dirichlet parameters $(\alpha,\beta)$. Accordingly, we can calculate $p(Y=y_1|x) = Dir(\alpha)$ and $p(Y=y_2|x) = Dir(\beta)$. 

To understand whether $p(Y=y|x)$ is high enough for $y$ to be considered consistent with $x$, one possible approach is to consider the \emph{likelihood ratio}, as considered in several works of Sampling-based Approximate Inference~\cite{lik}. This ratio is $\frac{p(Y=y|x)}{p(Y=y)}$. If this ratio is greater than 1, it means that the projected results are more likely than usual if conditioned on the actual result, which is an affirmation of the survey. On the other hand, the ratio being 1 or less suggest that the projected results may be dubious or independent of the actual results. 

However, calculating $p(Y=y)$ is computationally expensive as it involves marginalizing over both $Z$ and $X$. Unlike $p(Y/X)$, we cannot express $p(Y)$ as a Dirichlet distribution, as possible values of $Y$ and their respective probabilities are too varied to be expressed by a single distribution. A possible approach is the Dirichlet Process Mixture Model (DPMM) with Dirichlet base distribution, but even then, calculating the marginal likelihood is very difficult~\cite{dpmm}. So we carry out an alternate non-parametric approach based on Monte-Carlo Sampling, similar to the sampling procedure from $P(X|Y)$ as discussed in Sec 4.1.

\begin{enumerate}
    \item Draw $N$ candidate samples of vote share $\{X^c_{11},\dots,X^c_{1N}\}$ from the prior $g(X)$
    \item From each of them, sample an election $\{Z^c_{11},\dots,Z^c_{1N}\}$ using Election Model
    \item Simulate surveys on them using Survey Model and obtain projected vote shares $\{Y^c_{11},\dots,Y^c_{1N}\}$ and seat shares $\{Y^c_{21},\dots,Y^c_{2N}\}$ 
    \item Find the number of samples of $Y$ that are within a specified distance of both $y_1$ and $y_2$,
\end{enumerate}
So, the density at any arbitrary projection $y=\{y_1,y_2\}$ can be obtained as $p(y)\approx \frac{1}{N}\sum_{i=1}^NI(KL(y^c_{1i},y_1)<\epsilon_1)I(KL(y^c_{2i},y_2)<\epsilon_2)$.

Similarly, $p(y|X)$ is obtained in the same way, but by considering only those samples from $\{X^c_{11},\dots,X^c_{1N}\}$ for which are close enough to $X_1$, and the corresponding $\{X^c_{21},\dots,X^c_{2N}\}$ are also close enough to $X_2$. Closeness is once again measured in terms of K-L Divergence. We call the ratio $\frac{p(y|X)}{p(y)}$ thus obtained as the \textbf{nonparametric likelihood ratio}.

An alternate approach is to calculate $\frac{p(Y=y|x)}{max_{Y}p(Y|x)}$, i.e. how likely are the projected results compared to the most likely projections from an ideal survey. The denominator can be easily calculated using the estimated Dirichlet parameters of $p(Y|x)$. We call this ratio as the \textbf{likelihood mode ratio}.

\section{Experimental Evaluation}
In this section, we discuss detailed validation of the concepts discussed above on simulated data, and then proceed to evaluate actual political elections and surveys held in India. The main questions we wish to validate here are as follows: i) Does the Survey Model produce realistic results from an election? ii)  how does the accuracy of a survey depend on its scale? iii) Can the constructed Dirichlet Posterior $q(Y|Z)$ distinguish between fair and biased surveys? iv) Can we predict the election results from fair surveys using the constructed posterior? v) Can we estimate the performance of a party based on fair surveys? vi) Can we evaluate actual political elections using this setting? Below, we describe detailed experiments to answer the questions.

\subsection{Survey Model Evaluation}
While a single survey's result $Y$ is stochastic (depending on the sample of respondents and districts chosen), we can construct the distribution over projected results by Monte Carlo sampling using the Survey Model. To understand this, we construct a small experiment over $N=10000$ electors, $S=5$ districts and $K=3$ parties. These 5 seats can be divided among the 3 parties in 21 ways ((5,0,0), (2,3,0), (1,1,3) etc). We consider two different vote-shares: $(0.4,0.35,0.25)$ and $(0.5,0.4,0.1)$ over the 3 parties. In the first case there is close contest, while in the second case there is a prominent winner and loser. However, these votes may be distributed across the districts in different ways, resulting in different seat shares - from (2,2,1) to (5,0,0). The question is, can the survey results reflect these? Are the modes of the survey distributions located at these outcomes? If not, how far from the modes are they?

The complete election results $Z_1$ and $Z_2$ (corresponding to these two vote-shares) are generated using the DPM/SPM Election Model with concentration parameter 0.9. The seat shares obtained are $(2,2,1)$ and $(3,1,1)$ respectively. 

The Survey Model is then applied on both $Z_1$ and $Z_2$ 1000 times, and the projected seat shares are recorded in each case. We consider $f_n=0.1$ and $f_s=1$ (i.e. $10\%$ people are queried from all of the districts uniformly). Thus, we obtain empirical frequency distributions over the 21 possible seat distributions. It is found that for $Z_1$, the \emph{accurate seat projection rate} is $65.4\%$, i.e. the projected seat share matches the true seat share $65.4\%$ times. Other results which had significant probability under the survey were $(3,1,1)$ and $(3,2,0)$, both close to the correct result. For $Z_2$, this figure is $53.7\%$. 

We scale up the experiments to $N=1000000, S=100$ and repeat for other values of the concentration parameter of SPM. The $(f_n,f_s)$ parameters are held at $(0.1,1)$. The accurate seat projection rate for both vote shares and different concentration values are shown in Fig. 1, for different margins of error (for example, if the true seat distribution is $(50,30,20)$ and projected one is $(51,28,21)$ we can say that error is within 2). It is observed that in case of Fig 1a (comparable vote shares), performance is better for higher values of concentration, i.e. when the seat share reflects the vote share more closely. In case of Fig 1b (diverse vote shares), the relation is less clear. But the accurate seat projection rate is significantly higher compared to Fig 1a. This means, when the election is closely contested in terms of vote share, surveys are more likely to be accurate if the seat shares are compatible with vote shares. In case of lopsided elections in terms of vote share, surveys are generally expected to be more accurate.
\begin{figure}
    \centering
    \includegraphics[width=6cm,height=6cm]{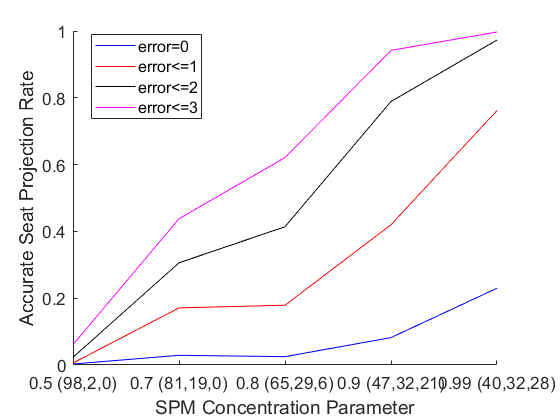}\includegraphics[width=6cm,height=6cm]{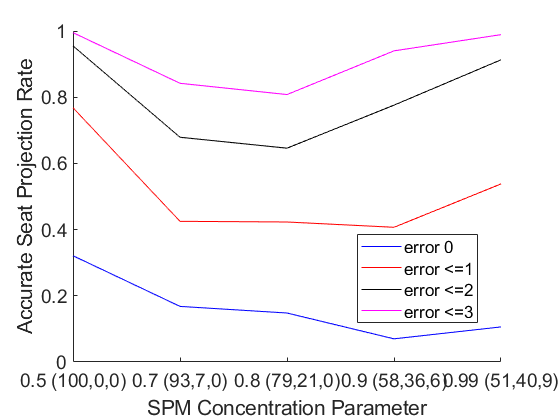}
    \caption{Comparison of Accurate Seat Projection Rate for different vote and seat shares. Fig 1a (left): close contest with vote shares (0.4, 0.35, 0.25), Fig 1b (right): lopsided contest with vote shares (0.5, 0.4, 0.1)}
    \label{fig:1}
\end{figure}

Should a survey go wider (cover more districts) or deeper (ask more people in each district)? We study how the accurate seat projection rate varies with the scale of the survey, i.e. with $f_n$ and $f_s$. We repeat this experiment for both the aforementioned vote shares, and also the two SPM concentration parameters (0.9 and 0.7) resulting in different seat shares. High concentration causes the seat share to reasonably resemble the vote share, while low concentration maximizes seat share of the party with highest vote share. The results are shown in Fig. 2. In Fig 2a (left) the number of people surveyed is varied, while keeping the district coverage unchanged ($50\%$), while in Fig 2b (right) the district coverage is varied, while keeping the people coverage unchanged ($10\%$). 

It is observed that covering more people has no clear impact when the concentration is high, i.e. seat share reflects the vote share. But for low concentration, covering more people clearly improves the survey performance. On the other hand, covering more districts is clearly more beneficial in case of high concentration, but not so much in case of low concentration.

\begin{figure}
    \centering
    \includegraphics[width=6cm,height=6cm]{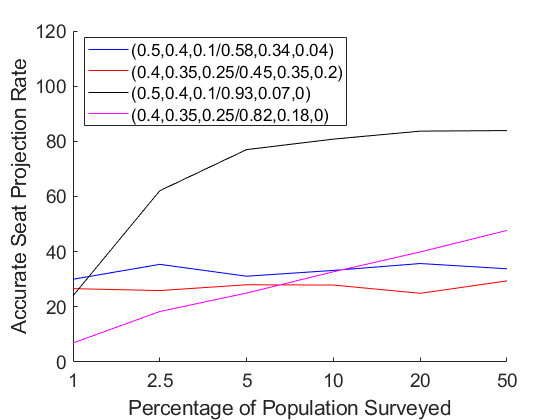}\includegraphics[width=6cm,height=6cm]{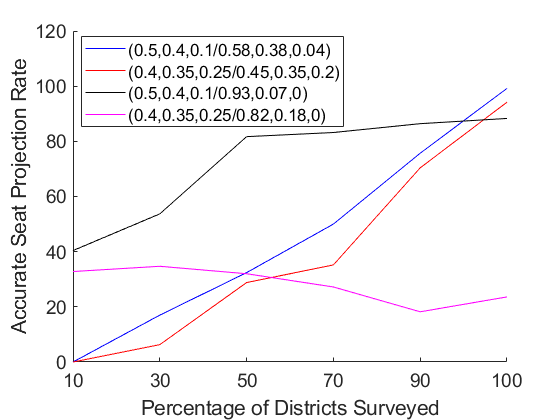}
    \caption{Comparison of Accurate Seat Projection Rate for surveying different fractions of the population (fig 2a-left), and districts (fig 2b-right) on 4 vote share/seat share combinations (see legend). Error limit: $3\%$}
    \label{fig:2}
\end{figure}

The above observations are validated further on actual political elections held in India. We consider four states of India (Tripura, Himachal Pradesh, Gujarat, Karnataka) that had elections in the past year. All of these were essentially tripartite contests, where the vote shares and seat shares of the three main parties are provided in Table 1. To avoid needless controversies, we have anonymized the parties. In each case, we refer to the party with most votes as P1, second most as P2 etc.
\begin{table}[]
    \centering
    \begin{tabular}{|c|c|c|c|c|c|c|c|c|}
    \hline
    State &N & S & \multicolumn{3}{|c|}{Vote Share} & \multicolumn{3}{|c|}{Seat Share}\\
    \hline 
    & & & P1 & P2 & P3 & P1 & P2 & P3\\
    \hline
    Tripura & 2.4M & 60 & 0.42 & 0.38 & 0.20 & 0.55 & 0.23 & 0.22\\
    Himachal & 4.2M & 68 & 0.45 & 0.43 & 0.12 & 0.59 & 0.37 & 0.04\\
    Gujarat & 29M & 182 & 0.56 & 0.30 & 0.14 & 0.88 & 0.09 & 0.03\\
    Karnataka & 36M & 224 & 0.46 & 0.39 & 0.15 & 0.62 & 0.30 & 0.08\\
    \hline 
    \end{tabular}
    \caption{Summary of 4 recent state assembly elections in India. Parties anonymized and ranked in order of vote share}
    \label{tab:my_label}
\end{table}
Surveys are simulated by the Survey Model using the complete election data obtained from ~\cite{eci}. Once again we vary $f_n$ and $f_s$ as above, though $f_n$ is now kept to smaller values ($0.1-5\%$ of the total population) due to the huge sizes of the electorate. The results shown in Fig 3. In most cases, we see that increasing the district coverage results in clear improvement of projections ($f_n$ constant at $10\%$), while increasing people coverage has no such effect (district coverage held constant at $50\%$). This is consistent with our previous analysis, as in all cases (except Gujarat) the seat shares are not very far from the vote share. The optimal SPM concentration parameter in all these cases, using which the seat share can be obtained most accurately given the vote shares, is found to be around 0.9. So this observation is consistent with our previous result (Fig 2).

\begin{figure}
    \centering
    \includegraphics[width=6cm,height=6cm]{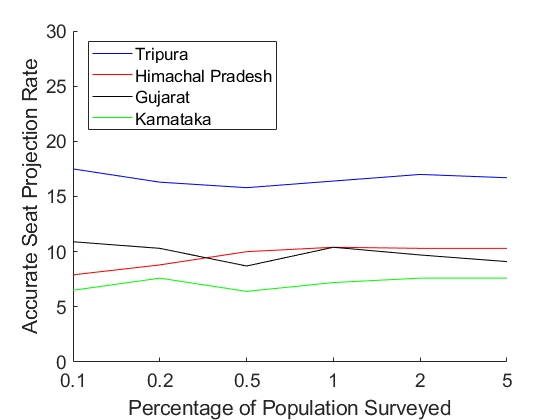}\includegraphics[width=6cm,height=6cm]{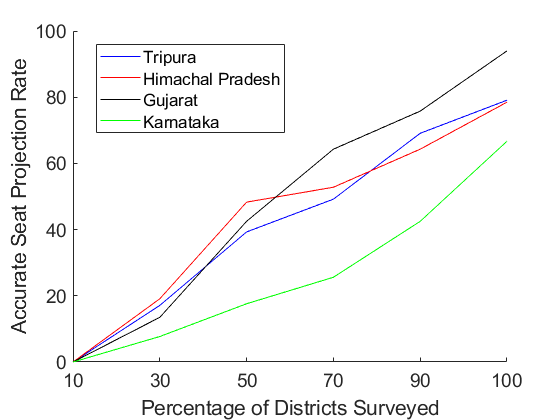}
    \caption{Comparison of Accurate Seat Projection Rate for surveying different fractions of the population (fig 2a-left), and districts of Indian state elections. Maximum error: $3\%$}
    \label{fig:3}
\end{figure}

\section{Predicting Results from Surveys}
We now set out to evaluate the constructed posterior $p(X|Y)$, i.e. given the survey projections, how well can we identify which outcomes are most likely, and which are not? For this, we carry out three experiments.

In the first experiment, we consider the true result $X^0=\{X^0_1,X^0_2\}$ and generate complete results from the election model. $X^0_1$ is obtained from the Dirichlet prior $r$. The survey model is run on it to generate a projection $\{Y_1,Y_2\}$, considering $N=1000000,S=100$. Now, we develop the posterior, by Monte Carlo Sampling and Maximum Likelihood estimate of Synthetic Dirichlet parameters as discussed in Sec 3.1. We now calculate the posterior density at a number of candidate results, including $X^0$. This is repeated for three sets of results: i) $X^0_1=(0.55,0.23,0.22)$, $X^0_2=(0.72,0.15,0.13)$, ii) $X^0_1=(0.35,0.33,0.32)$, $X^0_2=(0.36,0.36,0.28)$, iii) $X^0_1=(0.35,0.33,0.32)$, $X^0_2=(0.71,0.27,0.02)$. Note that ii) and iii) have identical vote shares but very different seat shares (due to different values of SPM concentration). Among the candidate solutions in each case, $X^0$ and results closest to it are the ones with highest posterior density. Fig. 4 shows how the posterior density at different results decreases as their distances (K-L Divergence) from the original result $X^0$ increases. Table 2 shows the true results $X^0$, projected results $Y$ and candidate solution with highest posterior density. 

\begin{table}[]
    \centering
    \begin{tabular}{|c|c|c|}
    \hline
    Actual Results & Projections & Posterior Mode\\
    \hline
         $(0.55,0.23,0.22)$ & $(0.51,0.26,0.23)$ & $(0.55,0.23,0.22)$\\
         $(0.35,0.33,0.32)$ & $(0.34,0.33,0.33)$ & $(0.34,0.32,0.33)$\\
         $(0.35,0.33,0.32)$ & $(0.36,0.34,0.30)$ & $(0.35,0.33,0.32)$\\
    \hline
    \end{tabular}
        \begin{tabular}{|c|c|c|}
    \hline
         $(0.72,0.15,0.13)$ & $(0.76,0.14,0.10)$ & $(0.72,0.15,0.13)$\\
         $(0.36,0.36,0.28)$ & $(0.36,0.34,0.30)$ & $(0.37,0.30,0.33)$\\
         $(0.71,0.27,0.02)$ & $(0.54,0.34,0.12)$ & $(0.71,0.27,0.02)$\\
    \hline
    \end{tabular}
    \caption{Original, projected, posterior mode vote shares (above) and seat shares (below) for three candidate settings}
    \label{tab:2}
\end{table}

\begin{figure}
    \centering
    \includegraphics[width=6cm,height=6cm]{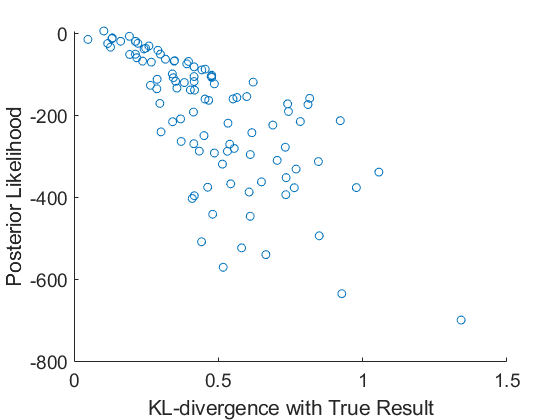}\includegraphics[width=6cm,height=6cm]{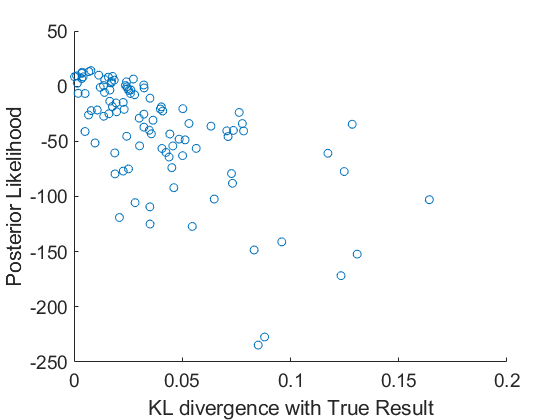}
    \caption{Relation between posterior density of candidate solutions and their distance (K-L divergence) from the actual result $X^0$. 4a(left): $X^0=(0.55,0.23,0.22)/(0.72,0.15,0.13)$, 4b(right): $X^0=(0.35,0.33,0.32)/(0.36,0.36,0.28)$}
    \label{fig:4}
\end{figure}

Note that for case ii) the highest posterior density value is achieved at $X_1=(0.34,0.33,0.33)$, $X_2=(0.37,0.33,0.30)$ which is different from, but very close to $X^0$. In cases i) and iii) $X^0$ has the best posterior density.

How does the posterior's performance change with the number and scales of the survey? This is the question we study in the third experiment. We repeat the second experiment by varying the number of surveys, as well as $f_n$ and $f_s$ in each survey. We see that as we increase the number of surveys, the posterior density of the actual result increases with respect to other candidates. For example, in case of setting ii) above, $X^0$ has the highest posterior density when we consider 5 surveys (which was not the case when we considered 1 survey). The results are shown in Appendix.

In the second experiment, we consider the election results from the four state elections discussed in Table 1. We consider 5 surveys in each case, by using our survey model on the complete election data. Next, the posterior density is computed for several candidate solutions including $X^0$. Once again, Fig. 5 shows how the posterior density at different candidate results vary with their distances from the actual results.

\begin{figure}
    \centering
    \includegraphics[width=6cm,height=6cm]{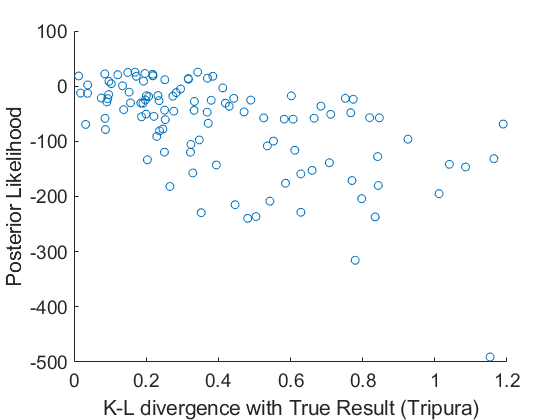}\includegraphics[width=6cm,height=6cm]{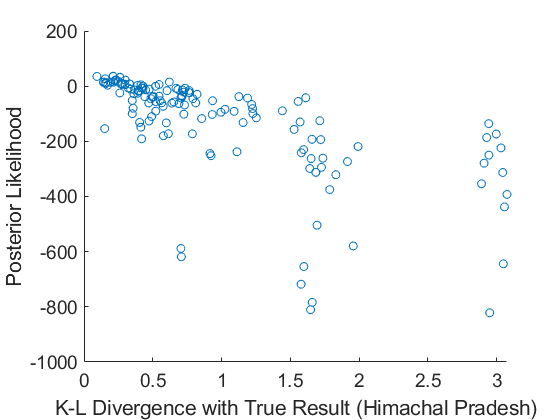}\\
    \includegraphics[width=6cm,height=6cm]{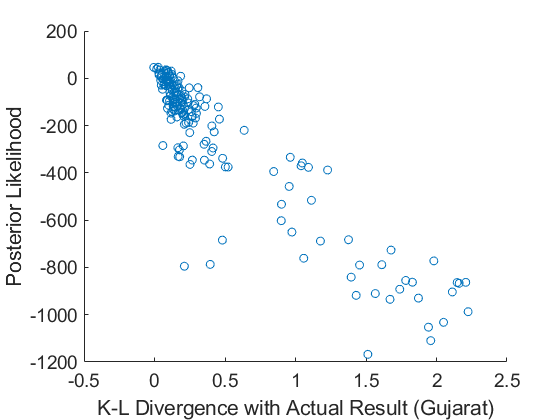}\includegraphics[width=6cm,height=6cm]{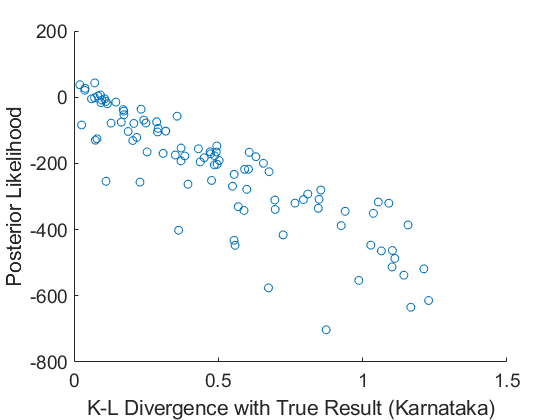}
    \caption{Posterior Likelihood of candidate results versus their distance (K-L divergence) from the actual results in case of the 4 state elections. Note the anomalous nature of the plot for Tripura, which had a multi-modal posterior}
    \label{fig:5}
\end{figure}

In case of Tripura, the SPM model fails to produce the true results under any parameter settings. So we consider the PCM model. Even then, the few candidate results with highest likelihood were quite varied: including $(0.39,0.36,0.25)/(0.55,0.4,0.05)$ and $(0.41,0.37,0.22)/(0.9,0.1,0)$. In both cases, either the vote-share or the seat-share are reasonably close to the actual, but not both.  This is a special case of a \emph{multi-modal posterior}, where varied results seem to be equally likely. This is reflected in the nature of the plot in Fig 5. The reason is that, $P3$'s vote-share was extremely skewed across districts. In case of Himachal Pradesh, the most likely result according to SPM model, based on 5 surveys is $(0.46,0.43,0.11)/(0.6 0.4 0)$. This result has a slightly higher posterior likelihood than the actual result. The SPM model was generally unable to produce results that allocate $0.14$ seat share to P3. In case of Gujarat and Karnataka, the actual result itself had the best likelihood among the candidate results which we considered. The comparisons of the actual result, projected results (median from 5 surveys) and posterior mode results are provided in Table 3, except for Tripura where there is no clear posterior mode. The conclusion is that, the constructed likelihood is consistent, i.e. it is able to recover the true result from the surveys in most cases. In the Appendix, we show how these results change with the number and scale of surveys, and the prior distribution $g(X_1)$. 

\begin{table}[]
    \centering
    \begin{tabular}{|c|c|c|c|}
    \hline
    State & Actual Results & Projections & Posterior Mode\\
    \hline
     Himachal&   $(0.45,0.43,0.12)$ & $(0.44,0.45,0.11)$ & $(0.46,0.43,0.11)$\\
     Gujarat&   $(0.56,0.30,0.14)$ & $(0.57,0.27,0.13)$ & $(0.56,0.30,0.14)$\\     
     Karnataka& $(0.46,0.39,0.15)$ & $(0.47,0.38,0.15)$ & $(0.46,0.39,0.15)$\\  
    \hline
    \end{tabular}
        \begin{tabular}{|c|c|c|c|}
    \hline
        Himachal&   $(0.59,0.37,0.04)$ & $(0.53,0.40,0.07)$ & $(0.54,0.44,0.02)$\\
        Gujarat&   $(0.88,0.09,0.03)$ & $(0.87,0.09,0.03)$ & $(0.88,0.09,0.03)$\\  
        Karnataka& $(0.61,0.30,0.09)$ & $(0.62,0.28,0.10)$ & $(0.61,0.30,0.09)$\\  
    \hline
    \end{tabular}
    \caption{Original, projected, posterior mode vote shares (above) and seat shares (below) for each party in the 3 state elections except Tripura. The projected results mentioned are based on the median of 5 surveys. with an error range of $\pm0.05$ around the median.}
    \label{tab:3}
\end{table}

\subsection{Party-specific Performance Distribution}

A related question that arises is, given survey $Y$, what can we say about the probable performance of a particular party? Our approach to this question has already been discussed in Section 4.1. We evaluate the same using the same 4 state elections as above, based on 5 surveys. Table 4 shows the modal results vis-a-vis survey results for the 4 states, for both vote-share and seat-share. We can see that the approximate posterior mode is quite accurate for vote share, but not very accurate in terms of seat share. In Fig 6, we plot the synthetic posterior PDF for the first, second and third parties (both vote share and seat share) conditioned on the 5 survey results for the elections. We find that in each case, the modes for the parties' curves are in the correct order of their actual performance, though there are significant variances, which means there is some probability that the results may have been different. For Tripura, the variances are very small and modes very close, while for Gujarat and Karnataka the seat share variance is quite large for $P1$. 

\begin{figure}
    \centering
    \includegraphics[width=6cm,height=5cm]{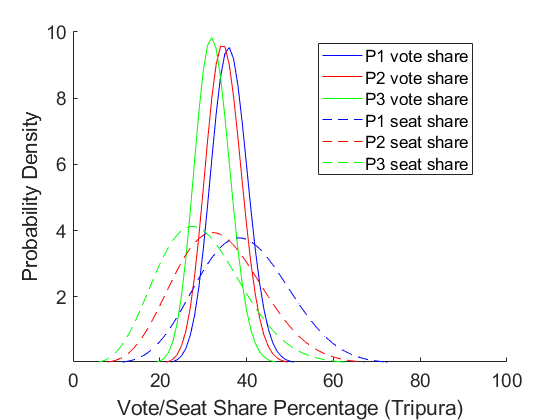}\includegraphics[width=6cm,height=5cm]{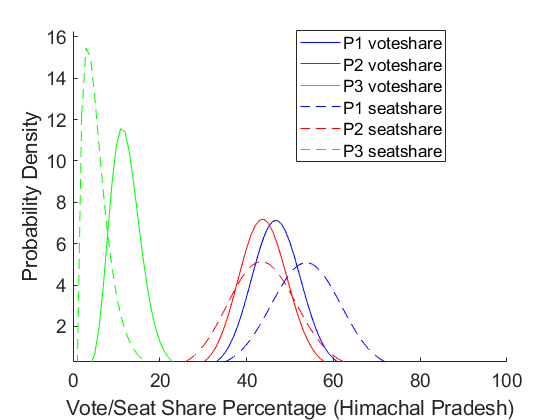}\\
    \includegraphics[width=6cm,height=5cm]{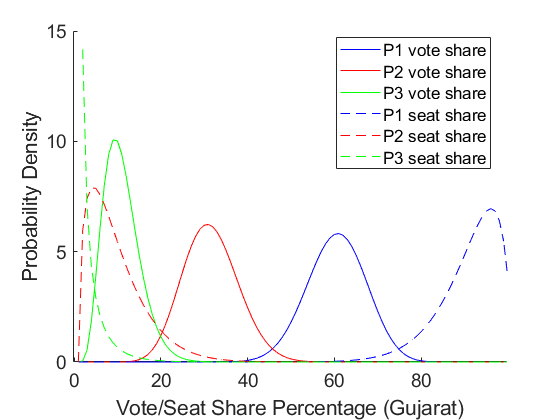}\includegraphics[width=6cm,height=5cm]{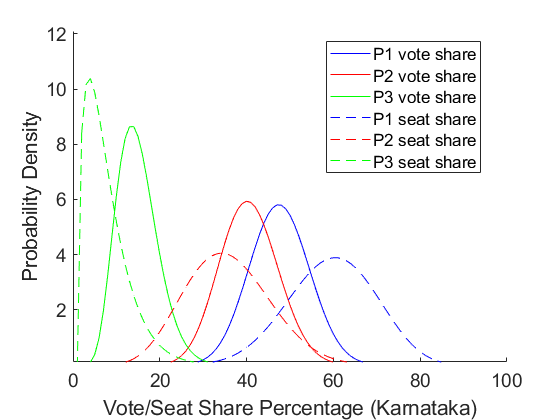}
    \caption{Synthetic Posterior distributions for the performance of each party individually, in terms of vote share and seat share, conditioned on 5 surveys for the 4 state elections.}
    \label{fig:6}
\end{figure}

\section{Impact of Survey Settings on Posterior}
The two key contributions of this paper are the survey model and the synthetic posterior likelihood. In Sections 5.1 and 5.2 we have seen some experimental validation of these, for both synthetic and real election data. The main parameters of the survey model are $n$: fraction of the population that is covered, and $s$: fraction of the districts that is covered. The synthetic likelihood construction process utilizes both of these parameters in simulating the surveys. In addition, it involves two more factors: i) the number of surveys available, and ii) the choice of the prior distribution $g(X_1)$. Here, we discuss the roles of these parameters and factors in greater details. Figures 2 and 3 of the main paper show the impact of $n$ and $s$ on the performance of the survey model. But here we shall see, how they impact the posterior distribution.  

\subsection{Prior distribution}
First of all, we consider the prior distribution $g(X_1)$. Clearly this is a Dirichlet Distribution with hyperparameters $(\gamma_1,\dots,\gamma_K)$. These hyperparameters may indicate our prior belief on the relative performance of the different parties, maybe based on the past election performances. The prior mode is then $\{\frac{\gamma_1-1}{\sum_k\gamma_k-K},\dots,\frac{\gamma_K-1}{\sum_k\gamma_k-K}\}$.
But if a survey $Y$ provides results which are significantly different from the prior mode, then there is a contradiction. This can easily happen if an election has significantly different, even reverse results compared to the previous one. In such a case, for any candidate outcome $X$, the prior $r(X)$ and the likelihood $q(Y|X)$ can be contradictory (one high, the other low). To understand what happens to the posterior in that case, we carried out simulation studies for the Himachal Pradesh state elections (details in Table 1, main paper). The survey settings were kept constant at $(n=0.01,s=0.25)$ and 5 survey was considered. 5 different values of prior hyperparematers $\gamma$ were considered, and synthetic posterior density was estimated for 100 candidate solutions (including the actual one) in each case. 

Figure 1 shows the relation between the posterior density of various candidate results and their K-L Divergence with the true result (Table 1, main paper). This is done using all 5 prior hyperparameter settings. In general, we desire that the candidate outcomes more similar to the actual outcome should have higher posterior density. Figure 1 shows that this property generally holds for all prior settings. However, if we consider only 1 survey (instead of 5), then the results become more sensitive to the prior hyper-parameters. In fact, we find that many candidate outcomes with varying similarities with the actual results have comparable posterior density, indicating a multi-modal posterior. This is demonstrated in Table 2, which shows the top candidate outcomes in each prior hyperparameter setting based on 1 survey. We find that this top candidate result is somewhat different from the actual result in each case. However, we also see in Table 2 that the density at the actual result also changes with prior hyperparameters, and it is maximum when the hyperparameters for top-2 parties are equal.

\begin{figure}
    \centering
    \includegraphics[width=9cm,height=6cm]{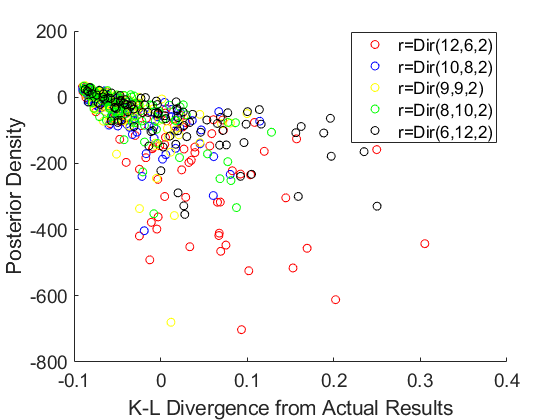}
    \caption{Posterior Likelihood of candidate results versus their distance (K-L divergence) from the actual results in case of the 5 prior hyperparameter settings. }
    \label{fig:8}
\end{figure}

\subsection{Number of Surveys}

\begin{figure}
    \centering
    \includegraphics[width=6cm,height=6cm]{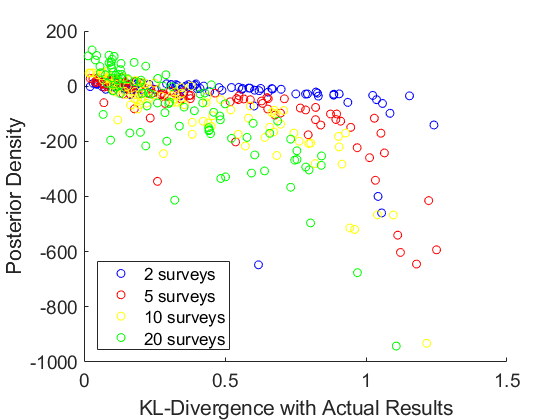}\includegraphics[width=6cm,height=6cm]{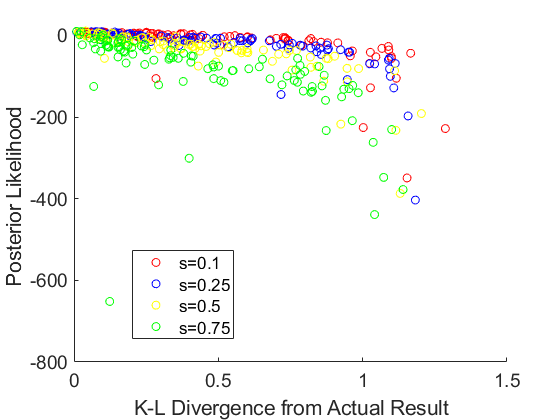}
    \caption{Impact of increasing the number of surveys on the posterior likelihood (left), Impact of increasing the fraction $s$ of districts covered by surveys on the posterior likelihood (right).}
    \label{fig:9}
\end{figure}

The number of surveys based on which we carry out the posterior inference, is another important factor on which the goodness of the inference depends. This is particularly true in case of very close elections. In such cases, different surveys can yield different or opposite results even if they cover reasonable number of districts and persons. We tested the case of Himachal Pradesh assembly election, where P1 had a small advantage over P2 in terms of vote share, but a big advantage in terms of seat share (Table 1 of main paper). We tested the posterior density obtained using 2,5,10 and 20 surveys, and the results are shown in Figure 2. We find that in case of smaller number of surveys, many candidate outcomes have high posterior density though they are quite far (in terms of KL-Divergence) from the actual outcomes. This is confirmed by the top-10 candidate outcomes shown in Table 3 (top), which includes many cases where P2 outperforms P1. However, this situation improves when we consider more surveys, as the candidate outcomes that are far from the actual results have lesser and lesser posterior density (Fig 8). The top-10 candidate outcomes for 20 surveys, as shown in Table 3 (bottom), also shows many cases which are close to the actual results.

\begin{table}[]
    \centering
    \begin{tabular}{|c|c|c|c|}
    \hline
        Hyper- & Survey Result & Most likely & Actual Result\\
        parameters & & outcome & Density\\
        \hline  
        (6,12,2) & (0.48,0.42,0.1) & (0.46,0.43,0.11) & 6.41\\
        & (0.76,0.16,0.08) & (0.59,0.4,0.01) &\\
        \hline
        (8,10,2) & (0.48,0.42,0.1) & (0.49,0.41,0.1) & 7.29\\
        & (0.76,0.16,0.08) & (0.72,0.28,0) &\\
        \hline
        (9,9,2) & (0.48,0.42,0.1) & (0.48,0.42,0.1) & 7.5\\
        & (0.76,0.16,0.08) & (0.63,0.37,0) &\\  
        \hline
        (10,8,2) & (0.48,0.42,0.1) & (0.51,0.4,0.09) & 7.48\\
        & (0.76,0.16,0.08) & (0.65,0.31,0.04) &\\  
        \hline
        (12,6,2) & (0.48,0.42,0.1) & (0.49,0.42,0.09) & 6.5\\
        & (0.76,0.16,0.08) & (0.78,0.22,0) &\\     
        \hline
    \end{tabular}
    \caption{Impact of Prior Hyperparameters on Posterior Density of voteshare (above) and seat share (below). True result: $(0.45,0.43,0.12)/(0.59,0.37,0.04)$}
    \label{tab:2}
\end{table}

\begin{table}[]
    \centering
    \begin{tabular}{|c|c|c|c|c|c||c|c|c|c|c|c|}
    \hline
    \multicolumn{3}{|c|}{Vote Share} & \multicolumn{3}{|c||}{Seat Share} & \multicolumn{3}{|c|}{Vote Share} & \multicolumn{3}{|c|}{Seat Share} \\ \hline
    \hline
        P1 & P2 & P3 & P1 & P2 & P3 & P1 & P2 & P3 & P1 & P2 & P3  \\ \hline
        0.44 & 0.47 & 0.09 & 0.38 & 0.54 & 0.07 & 0.44 & 0.45 & 0.11 & 0.46 & 0.44 & 0.1 \\ \hline
        0.45 & 0.46 & 0.09 & 0.37 & 0.56 & 0.07 & 0.46 & 0.44 & 0.1 & 0.57 & 0.43 & 0 \\ \hline
        0.41 & 0.47 & 0.12 & 0.4 & 0.49 & 0.12 & 0.4 & 0.45 & 0.15 & 0.44 & 0.47 & 0.09\\ \hline
        0.4 & 0.51 & 0.08 & 0.37 & 0.57 & 0.06 & 0.47 & 0.39 & 0.13 & 0.65 & 0.28 & 0.07 \\ \hline
        0.42 & 0.43 & 0.15 & 0.44 & 0.47 & 0.09 & 0.41 & 0.49 & 0.1 & 0.38 & 0.56 & 0.06 \\ \hline
        0.41 & 0.44 & 0.14 & 0.46 & 0.49 & 0.06 & 0.47 & 0.46 & 0.08 & 0.56 & 0.44 & 0 \\ \hline
        0.44 & 0.41 & 0.14 & 0.51 & 0.44 & 0.04 & 0.53 & 0.37 & 0.1 & 0.68 & 0.28 & 0.04\\ \hline
        0.48 & 0.44 & 0.07 & 0.57 & 0.4 & 0.03 & 0.47 & 0.45 & 0.08 & 0.57 & 0.43 & 0\\ \hline
        0.48 & 0.45 & 0.06 & 0.59 & 0.38 & 0.03 & 0.42 & 0.46 & 0.11 & 0.37 & 0.62 & 0.01 \\ \hline
        0.41 & 0.53 & 0.06 & 0.41 & 0.59 & 0 & 0.48 & 0.43 & 0.08 & 0.65 & 0.35 & 0  \\ \hline
    \end{tabular}
    \caption{Top-10 most likely candidate outcomes using 2 (left) or 20 (right) surveys for Himachal Pradesh Assembly Election, with $n=0.01,s=0.25$}
    \label{tab:3}
\end{table}

\subsection{District and Person Coverage}
Next, we we consider the number of districts covered in the survey. We vary this fraction $s$ as $\{0.1,0.25,0.5,0.75\}$, keeping the person coverage constant as $1\%$, with equal number of persons being queried in each of the chosen districts. The prior hyper-parameters are kept at $(8,10,2)$ and the number of surveys is 1. Our aim is to see how the posterior distribution of the candidate outcomes varies as the $s$ parameter is changed. We study this for the case of Himachal Pradesh, which had a very close election with P1 and P2 having very close vote shares. The results are shown in Figure 8, where we see when the district coverage is low, many candidate outcomes which are far from the actual outcome also have quite high posterior likelihood. However, such cases become rarer as we increase the district coverage. Finally, we consider the number of persons covered in the survey. We vary this fraction $n$ as $\{0.001,0.01,0.05,0.1\}$, keeping the district coverage constant as $0.25$, with equal number of persons being queried in each of the chosen districts. However, In Fig 9, we find that increasing $n$ does not have a very clear impact on the posterior density, in case of Himachal Pradesh where the election was very close. But for Karnataka, where the winning margin for P1 was larger, the impact of increasing person coverage seems to make a clearer impact, as candidate outcomes further away from the actual results have less posterior density.

\section{Baseline Methods}
We have not provided any comparisons of our proposed approach in the main paper, because there are no known works with the same aim as ours. However, if our aim is restricted to obtain a distribution over the possible outcomes (vote share and seat share), one possibility is to use Bayesian prior-posterior analysis using Dirichlet distributions. We may consider two separate Dirichlet priors for $X_1$ and $X_2$, i.e. $X_1\sim Dir(\alpha_0), X_2\sim Dir(\beta_0)$. The survey results, i.e. the total number of responses $Y_1=(\{Y_{11},\dots,Y_{1K}\})*N*n$ obtained for the different parties, and the number of seats dominated by each party based on the responses $Y_2=(\{Y_{21},\dots,Y_{2K}\})*S*s$, are both considered to follow Multinomial Distribution. It is well-known that Dirichlet distribution is a Conjugate Prior for Multinomial, and so $X_1|Y_1\sim Dir(\alpha_0+Y_1), X_2|Y_2\sim Dir(\beta_0+Y_2)$. 

\begin{figure}
    \centering
    \includegraphics[width=6cm,height=6cm]{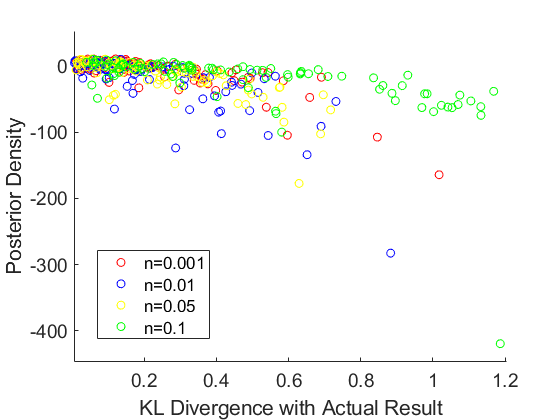}\includegraphics[width=6cm,height=6cm]{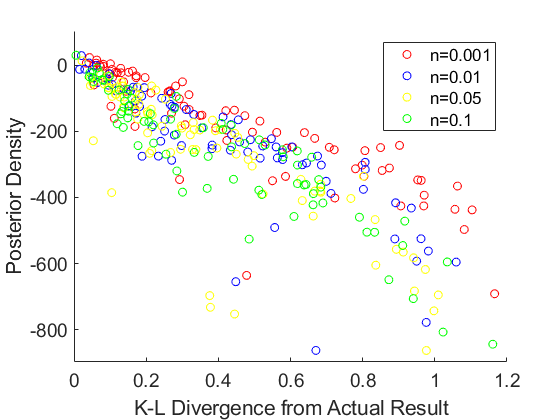}
    \caption{Impact of increasing the fraction $n$ of population covered by surveys on the posterior likelihood in case of Himachal Pradesh (left) and Karnataka (right).}
    \label{fig:10}
\end{figure}

The problem with this approach is that the vote-share and seat-share are considered independent of each other, which clearly they are not. Consequently, a candidate result $X$ where the $X_1$ and $X_2$ components are not compatible each other, i.e. the seat share $X_2$ is impossible given the vote share $X_1$, can still have a high posterior density as $X_1|Y_1$ and $X_2|Y_2$ can still be reasonably high individually. Furthermore, the density of $X_1$ tends to be numerically much larger than that of $X_2$, and hence a candidate result can have high posterior density based only on $X_1$, even if the density $X_2|Y_2$ is low. We observe this effect in case of the Tripura Assembly Election (Table 1), where the top-10 (highest posterior density) candidate results obtained by the proposed approach of Synthetic Likelihood are compared with the top-10 candidate results obtained from this Dirichlet-Multinomial baseline. We see that in the latter case, pretty much all the results have 0 seats for P3, whereas in reality it had 13 out of 60 seats. 

One way to offset this situation is by considering scaling factors on the Multinomial likelihood model, so that the posterior density values of $X_1$ and $X_2$ are comparable. This, however, introduces the reverse problem - there are many candidate solutions where P2 or P3 have a very large number of seats or vote share. 

Similar results were obtained in case of the other states (Himachal Pradesh, Gujarat, Karnataka) also. These results emphasize the fact that it is extremely important to include $X_1$ as a condition for the distribution of $X_2|Y_2$. But the relation between $X_1$ and $X_2$ is also not possible to express using any single distribution, due to which we must use a simulation-based approach as done in this work.

\begin{table}[!ht]
    \centering
    \begin{tabular}{|c|c|c|c|c|c|c|}
    \hline
        \multicolumn{3}{|c|}{Vote Share} & \multicolumn{3}{|c|}{Seats} &\\ 
        \hline
        P1 & P2 & P3  & P1 & P2 & P3  & Log-Density \\ \hline
        0.41 & 0.36 & 0.22 & 55 & 4 & 1 & 25.44 \\ \hline
        0.41 & 0.39 & 0.20 & 26 & 29 & 5 & 25.34 \\ \hline
        0.38 & 0.36 & 0.25 & 33 & 24 & 3 & 24.86 \\ \hline
        0.48 & 0.34 & 0.18 & 38 & 21 & 1 & 23.12 \\ \hline
        0.48 & 0.30 & 0.22 & 42 & 13 & 5 & 22.29 \\ \hline
        0.43 & 0.36 & 0.21 & 27 & 30 & 3 & 21.62 \\ \hline
        0.49 & 0.33 & 0.18 & 33 & 23 & 4 & 20.7  \\ \hline
        0.38 & 0.37 & 0.25 & 19 & 32 & 9 & 19.02 \\ \hline
        0.43 & 0.33 & 0.24 & 48 & 11 & 1 & 18.78 \\ \hline
        0.43 & 0.32 & 0.25 & 35 & 13 & 12 & 18.47\\ \hline
    \end{tabular}
    \begin{tabular}{|c|c|c|c|c|c|c|}
    \hline
        \multicolumn{3}{|c|}{Vote Share} & \multicolumn{3}{|c|}{Seats} &\\ 
        \hline
        \hline
        P1 & P2 & P3  & P1 & P2 & P3  & Log-Density  \\ \hline
        0.43 & 0.38 & 0.19 & 45 & 15 & 0 & -103.35  \\ \hline
        0.41 & 0.39 & 0.20 & 27 & 33 & 0 & -111.61  \\ \hline
        0.42 & 0.40 & 0.18 & 43 & 17 & 0 & -203.19  \\ \hline
        0.40 & 0.41 & 0.192 & 42 & 18 & 0 & -226.23  \\ \hline
        0.44 & 0.36 & 0.20 & 36 & 24 & 0 & -231.01  \\ \hline
        0.42 & 0.40 & 0.18 & 43 & 17 & 0 & -278.79  \\ \hline
        0.42 & 0.40 & 0.18 & 30 & 30 & 0 & -309.60  \\ \hline
        0.45 & 0.35 & 0.20 & 49 & 10 & 1 & -316.03  \\ \hline
        0.44 & 0.35 & 0.21 & 57 & 1 & 2 & -316.24  \\ \hline
        0.45 & 0.36 & 0.20 & 60 & 0 & 0 & -325.54  \\ \hline
    \end{tabular}
    \caption{Top-10 candidate outcomes (with posterior density) obtained from the proposed model (above) and Dirichlet-Multinomial baseline (below) in case of Tripura Assembly Election, based on 5 surveys covering $1\%$ of people and $25\%$ of the districts}
    \label{tab:1}
\end{table}

\section{Post-facto Survey Evaluation}
We finally validate the analysis of Section 4.2, to examine the validity of surveys once the actual result of the election is known. We compare three kinds of surveys: genuine, fake and malicious. Genuine surveys $Y_{gen}$ are generated by running the survey model on the actual complete election data $Z$. For fake surveys, a first a fake election $Z_{fake}$ is generated by first sampling a vote share $X_{fake}$ from the prior distribution, and then applying an election model on it. The survey model is then applied to obtain $Y_{fake}$. In case of malicious surveys, the true result is intentionally skewed towards one party. $Y_{mal}$ is obtained by linearly combining $Y_{gen}$ with $Y_{k}$ where the entire vote is in favour of party $k$ (chosen randomly). 

\begin{table}[]
    \centering
    \begin{tabular}{|c|c|c|c|}
    \hline
      Actual     & Genuine & Fake & Malicious\\
      \hline
      (0.4,0.35,0.25)|(0.4,0.4,0.2)  & 5.0 & 0.08 & 3.5\\
      (0.4,0.35,0.25)|(0.8,0.2,0)  & 11.7 & 2.23 & 1.29\\
      \hline
      \hline
      (0.4,0.35,0.25)|(0.4,0.4,0.2)  & 0.23 & 0.11 & 0\\
      (0.4,0.35,0.25)|(0.8,0.2,0)  & 1.0 & 0 & 0\\
    \hline
    \end{tabular}
    \caption{Survey Evaluation on synthetic election results simulated by SPM Election Model. Top: Nonparametric Likelihood Ratio, Bottom: Likelihood Mode Ratio (average of 100 surveys in each category)}
    \label{tab:4}
\end{table}

We first consider the synthetic election with $N=10000$ voters, $S=5$ districts and $K=3$ parties. We consider two cases: one where $X_1=(0.4,0.35,0.25),X_2=(0.4,0.4,0.2)$ and another where $X_1=(0.4,0.35,0.25),X_2=(0.8,0,0.2)$. For the three categories of surveys (genuine, fake, malicious), we compare both the nonparametric posterior likelihood and the posterior modal likelihood. The results are shown in Table 4. We clearly see that in every case, the genuine surveys have a significantly higher likelihood ratio or posterior modal ratio compared to the other surveys. 

Next, we move to the real data. We sample 100 surveys from each of the above 3 categories, for each of the 4 states. In each case, we calculate both the nonparametric likelihood ratio and likelihood mode ratio as discussed in Section 4.2. The mean results are reported in Table 5. Once again, we find that the \emph{genuine} surveys have a very significantly higher likelihood ratio compared to the fake or malicious cases. In case of Tripura, even for the Genuine surveys, the modal ratio is quite low because, the actual results could not be simulated accurately by any of the election models.

\begin{table}[]
    \centering
    \begin{tabular}{|c|c|c|c|}
    \hline
      State     & Genuine & Fake & Malicious\\
      \hline
      Tripura   & 10.9 & 2.5 & 1.6\\
      Himachal  & 17.1 & 0.72 & 3.11\\
      Gujarat   & 4.62 & 1.82 & 3.54\\
      Karnataka & 30.3 & 1.42 & 6.66\\
    \hline
    \hline
      Tripura   & 0.07 & 0.02 & 0\\
      Himachal  & 0.41 & 0.00004 & 0.0012\\
      Gujarat   & 0.13 & 0 & 0.0004\\
      Karnataka & 0.05 & 0.002 & 0 \\
    \hline
    \end{tabular}
    \caption{Top: Nonparametric Likelihood Ratio, Bottom: Likelihood Mode Ratio (average of 100 surveys in each of the 3 categories)}
    \label{tab:5}
\end{table}

\section{Computational Complexity}
One of the concerns with the proposed approach is its computational complexity, since it is based on Monte Carlo Simulations. In particular, we need a large number of samples of elections and surveys to compute the Synthetic Dirichlet parameters. To calculate $p(Y|X)$ we use $M$ samples of complete elections $\{Z_1,\dots,Z_M\}$ (Eq 2 of main paper), which are obtained using the Election Model. From each of these samples, we again draw $L$ sample surveys to calculate the Dirichlet parameters $\alpha(Z_i),\beta(Z_i)$ (Eq 4 of main paper). If we consider $t_1$ as the time to generate one election $Z_i$, $t_2(L)$ as the time to generate $L$ sample surveys from $Z_i$ and $t_3$ as the time to estimate $\alpha(Z_i),\beta(Z_i)$ from the sample surveys and calculate the Dirichlet density at $Y$ using them, then the total time to compute $p(Y|X)$ for a given $(X,Y)$ is $M*t_1+M*(t_2(L)+t_3)$. 

In our implementation, we consider $M=200, L=100$. The election model complexity $t_1$ is linear in $N$ (total number of electors), but a lot of speedup is achievable by considering the electors in batches. This partly compromises the SPM/ECM model, which assign the district $Z_i$ for each elector $i$, by sampling from the distribution $p(Z_i|Z_1,\dots,Z_{i-1})$. The distribution changes for each elector $i$, based on the $(i-1)$ electors before them. But here we consider batches of size $b$, where we sample the districts for $b$ electors simultaneous using a Multinomial Distribution based on the assignments to the electors before them. This parallelization significantly reduces $t_1$. For $N=10 million, S=100, b=100$, we have $t_1=1.5 secs$, while $t_2=0.25 secs$ for $n=0.01$ on a Intel i5 core (10th generation) processor using 8GB RAM. $t_2$ scales linearly with $N*n$. $t_3$ turns out to be just $0.05$ seconds. So for the given configuration, calculating $p(Y|X)$ takes about 5 minutes for a given $(X,Y)$.

\section{Discussions and Conclusion}
While much of the past work on election prediction from surveys focuses on prediction of the winner, there has been relatively few works on predicting the number of seats or votes won by different parties in a multi-party, multi-district setting. This work actually provides a probability distribution on these, and also on the possible performance of individual parties. Furthermore, we provide a way to evaluate the feasibility of survey results, once the actual results are known. This approach can be very useful in bringing scientific accuracy in the process of large-scale opinion polling and in identifying fraudulent or dubious surveys. The unique feature of this work is that it performs extensive simulations based on actual elections involving millions of people. While much of the work presented here is based on Monte Carlo simulations and Approximate Bayesian Computing, our next aims will be to provide some theoretical guarantees regarding the actual results on the basis of surveys. We have not provided any comparison of our proposed method, since there is no known approach to achieve the same target. However, in the Appendix, we discuss what could have been possible alternatives, and their shortcomings.

\bibliography{survey_archive}
\bibliographystyle{acm}


\end{document}